\documentclass[aps,pra,twocolumn,showpacs,preprintnumbers,amsmath,amssymb,floatfix,superscriptaddress]{revtex4-1}
\usepackage[english]{babel}
\usepackage[utf8]{inputenc}
\usepackage{microtype}
\usepackage{graphicx}
\usepackage{amsmath}
\usepackage{amssymb}
\usepackage{amsfonts}
\usepackage{mathtools}
\usepackage{xspace}
\usepackage{mathrsfs}
\usepackage{siunitx}
\usepackage{slashed}
\usepackage[inline]{enumitem}
\usepackage[hidelinks]{hyperref}
\usepackage[capitalize]{cleveref}
\usepackage{bbold}
\usepackage{color}
\usepackage{hyperref}
\usepackage{multirow}
\usepackage{epstopdf}

\usepackage{enumitem}
\setlist[itemize]{noitemsep}

\def\beq{\begin{equation}}
\def\eeq{\end{equation}}
\newcommand{\dd}[2]{\frac{d^2 #1}{d #2^2}} 
\newcommand{\expect}[1]{\left< #1 \right>} 
\newcommand{\ket}[1]{\left| #1 \right>}
\newcommand{\bra}[1]{\left< #1 \right|}
\newcommand{\A}{\ensuremath{\text{A}}}
\newcommand{\I}{\ensuremath{\text{I}}}

\newcommand{\affA}{Physics and Materials Science Research Unit, University of Luxembourg, L-1511 Luxembourg, Luxembourg}
\newcommand{\affB}{Department of Physics and Astronomy, Aarhus University, DK-8000 Aarhus C, Denmark}

\newcommand{\affD}{Institute for Advanced Study, Tsinghua University, Beijing 100084, China}
\newcommand{\affE}{Aarhus Institute of Advanced Studies, Aarhus University, DK-8000 Aarhus C, Denmark}
\newcommand{\affF}{Zentrum f\"ur Optische Quantentechnologien, Fachbereich Physik, and Hamburg Center for Ultrafast Imaging, Universit\"at Hamburg, 22761 Hamburg, Germany}

\begin{document}

%----------------------------------------------------------------------------------------
%	TITLE 
%----------------------------------------------------------------------------------------

\title{
Ion-induced interactions in a Tomonaga-Luttinger liquid
}

%----------------------------------------------------------------------------------------
%	AUTHORS AND AFFILIATIONS
%----------------------------------------------------------------------------------------

\author{A.~B.~Michelsen}
\affiliation{\affA}
\affiliation{\affB}
\author{M.~Valiente}
\affiliation{\affD}
\author{N.~T.~Zinner}
\affiliation{\affB}
\affiliation{\affE}
\author{A. Negretti}
\affiliation{\affF}

%----------------------------------------------------------------------------------------
%	ABSTRACT
%----------------------------------------------------------------------------------------

\begin{abstract}
We investigate the physics of a Tomonaga-Luttinger liquid of spin-polarized fermions superimposed on an ion chain. This compound system features (attractive) long-
range interspecies interactions. By means of density matrix renormalization group techniques
we compute the Tomonaga-Luttinger-liquid parameter and speed of sound as a function of
the relative atom/ion density and the two quantum defect parameters, namely, the
even and odd short-range phases which characterize the short-range part of the atom-ion
polarization potential. The presence of ions is found to allow critical tuning of the
atom-atom interaction, and the properties of the system are found to depend significantly
on the short-range phases due to the atom-ion interaction. These latter dependencies can
be controlled, for instance, by manipulating the ions' internal state. This allows modification
of the static properties of the quantum liquid via external driving of the ionic impurities.
\end{abstract}

\maketitle

%----------------------------------------------------------------------------------------
%	ARTICLE CONTENTS
%----------------------------------------------------------------------------------------

%
% SECTION
%

\section{Introduction}

The quantum physics of one-dimensional (1D) interacting systems is rather peculiar as quantum fluctuations are strong and only collective excitations are possible, i.e. there are no single-particle excitations typical of Fermi liquids. Because of this, when the transverse degrees of freedom are frozen and a system acts as if one-dimensional, counterintuitive phenomena occur, such as fermionisation (bosonisation) of bosons (fermions)~\cite{GirardeauJMP60,GirardeauPRL06}, perfect ``collisional transparency" of particles~\cite{OlshaniiPRL98} (equivalent to zero crossing of the two-body coupling constant), enhanced inter-particle interactions in a ballistic expansion~\cite{LiebPR63}, and unusual cooling mechanisms~\cite{RauerPRL16,SchemmerPRL18}, to mention a few. While decades ago such manifestations were regarded as mere mathematical curiosities, the advent of degenerate atomic quantum gases has allowed the verification of such predictions, as the atomic confinement can be designed via optical laser fields~\cite{GRIMM200095} or, alternatively, magnetic field landscapes can be engineered by means of tailored configurations of current-carrying wires in atom chips~\cite{QIP:ACbook11}. The understanding of the fundamental underlying mechanisms behind such phenomenology is not only of academic interest, but also has important practical applications, as the progressive miniaturisation of electronic devices is such that, for instance, any quantitative description of transport in extremely reduced spatial dimensions and extremely low temperatures must be quantum mechanical. 

Very recently, experimental advances in bringing different atomic systems together to form a hybrid quantum system have opened new possibilities for quantum physics research~\cite{tomzaCold2019a}. For instance, Rydberg~\cite{SchmidtPRL16,CamargoPRL18,SchmidtPRA18}  or other neutral  impurities~\cite{SpethmannPRL12,CataniPRA12,CetinaPRL15,CetinaS16,Jorgensen2016,Hu2016,Hohmann2015,WideraPRL18}  in quantum 
gases allow us to study the dressing of the atomic impurities with gas excitations and of mediated interactions~\cite{ChenPRL18,ChenPRA18,KinnunenPRL18,CamachoPRX18,CamachoPRL18,mistakidis2018,DehkharghaniPRL18} as well as to utilize impurities to probe bath correlations and temperature~\cite{SherkunovPRA09,KollathPRA07,RodriguezPRB18,Mehboudi2018}. In addition, charged or dipolar impurities in degenerate atomic gases allow us to study polarons in the strong coupling regime~\cite{CasteelsJLTP11}, to quantum simulate Fr\"ohlich-type Hamiltonians~\cite{BissbortPRL13} as well as extended Hubbard~\cite{Pupillo2008,Ortner2009,NegrettiPRB14,Baier201} and lattice gauge theories~\cite{DehkharghaniPRA17}. Experiments with an ion immersed in a Bose-Einstein 
condensate~\cite{ZipkesNature10,SchmidPRL10,HartePRL12,Kleinbach2018,Engel2018,MeirPRL16,Meir2018}, and in a Fermi gas~\cite{RatschbacherPRL13,Furst2017,Joger2017,ewald19} have been realised in recent years, albeit not yet in the deep quantum regime of atom-ion collisions. Specifically low dimensional quantum physics with impurities exhibits a variety of unusual quantum phenomena. A few examples of this are: Bloch oscillations experienced by a moving impurity in a strongly correlated bosonic gas without the presence of an optical lattice potential~\cite{MeinertS17}, quantum flutters~\cite{Mathy2012} (namely injected supersonic impurities that never come to a full stop), so-called infrared-dominated dynamics~\cite{KantianPRL14} and clustering of impurities~\cite{DehkharghaniPRL18}.

\begin{figure}[tb]
\centering
\includegraphics[width=\columnwidth]{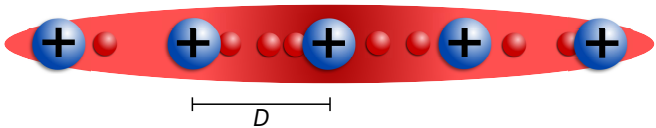}
\caption{Sketch of the physical system considered in this work. A linear ion crystal, whose ions are positively charged (big blue spheres) and separated by a distance $D$, and a Tomonaga-Luttinger liquid of ultracold atoms (indicated by the red cloud with small spheres) that overlaps the crystal. }
\label{fig:diagram}
\end{figure}

Motivated by these advances and by recent experiments that combine ytterbium ions with fermionic lithium atoms~\footnote{We note that currently the atom-ion species Li/Ca$^+$ is also under intense experimental investigations~\cite{HazePRL18}.}~\cite{Furst2017,Joger2017}, we investigate the ground state properties of a spin-polarised fermionic quantum gas that is superimposed on an ion chain (see Fig.~\ref{fig:diagram}), where the latter is treated statically. Given the fact that the motion of the ions and their internal states can be precisely controlled in experiments, atom-ion scattering properties can thus be manipulated. This can be useful e.g. for inducing macroscopic self-trapping or tunneling dynamics in a bosonic Josephson junction~\cite{GerritsmaPRL12,SchurerPRA16, ebghaCompound2019}. Here we are interested in the impact of the long-ranged atom-ion polarization potential on the 1D quantum fluid statical properties. Specifically, we employ density matrix renormalisation group techniques to extract the Tomonaga-Luttinger liquid (TLL) parameter and the speed of sound, which fully characterise the low energy physics of the atomic fluid. We find that these quantities have a significant dependence on the short-range physics of the atom-ion scattering (i.e., short-range phases), which can be controlled, for instance, by so-called confinement-induced~\cite{IdziaszekPRA07,MelezhikPRA16, melezhikImpact2019} or Fano-Feshbach resonances~\cite{IdziaszekPRA09,tomza15}. 
Thus, our findings demonstrate that the quantum fluid properties not only can be tuned by manipulating the ion quantum state, but also that this dependence is strong. 
As has been previously discussed, TLL's of 1D Bose-Fermi mixtures reveal a rich phase diagram~\cite{MatheyPRL04} and our goal is to understand how long-ranged interactions can affect the picture.

%
% SECTION
%

\section{Theoretical framework}

In this section we describe the system we study, the interaction between the two atomic species and between the fermionic atoms, as well as provide the basic ingredients of TLL theory that will be used later in the paper. 

\subsection{System Hamiltonian}

We consider an ensemble of identical ultracold atoms,
which are spin-polarised fermions,
confined to one spatial dimension in
the background of an ion chain with the ions organised as an evenly-spaced Coulomb crystal. The ions are considered
static, namely their motion is neglected, e.g., because of tight confinement or heavy ions and light atoms. 
We use realistic atom-ion interactions via an accurate mapping
of quantum defect theory (QDT) to an effective interaction potential
that also includes the asymptotic power-law tail
of the atom-ion forces. For the atom-atom interactions,
we use instead effective field theory (EFT), which is valid at low
energies and amenable to numerical treatment~\cite{valiente15}. The Hamiltonian for $N_\A$ atoms in the presence of an ion chain with $N_\I$ ions takes the form
\begin{align}
\label{eq:Hmicro}
\hat H &= \sum_{k=1}^{N_\A} \left[
\frac{\hat p_k^2}{2 m_A} + U(x_k) \right.
\nonumber\\
\phantom{=}&
+ \left. \sum_{j=1}^{N_\I} V_{\A\I}(x_k - X_j) + \sum_{j=1}^{N_\A} V_{\A\A}(x_k - x_j)
\right],
\end{align}
where $m_\A$ is the atom mass, $\hat p_k$ is the atomic momentum, $U(x)$ is the external trap (specifically a box-like potential), $V_{\A\I}(x_k - X_j)$ is the atom-ion interaction with $x_k$ and $X_j$ denoting the $k$'th atom position and the $j$'th ion position, respectively, and $V_{\A\A}(x_k - x_j)$ is the atom-atom interaction.
The atom-ion polarization potential is caused by the interaction between the ion electric field and the induced electric dipole of the atom. At long distances and in quasi 1D, it can be shown that 
the interaction takes the form~\cite{IdziaszekPRA07}
\begin{align}
V_{AI}(x - X) = - \frac{\alpha e^2}{2 (x - X)^4},
\label{eq:VAI}
\end{align}
where $e$ is the electron charge and $\alpha$ is the static polarisability of the atom. The potential, which is attractive and supports ion-bound atom states, is characterised by a characteristic length $R^*$ and energy $E^*$
\begin{align}
R^* = \sqrt{\frac{\alpha e^2 \mu}{\hbar^2}}, \qquad 
E^* = \frac{\hbar^2}{2 \mu (R^*)^2},
\label{eq:units}
\end{align}
where $\mu$ is the reduced atom-ion mass. 
Hereafter, all lengths are rescaled with respect to $R^*$.
As we already pointed out, we focus on the static ion scenario and a very favourable choice for the atom-ion pair is $^6\text{Li}$ / $^{174}\text{Yb}^+$. This pair appears to be the most promising to attain the ultracold regime in radio-frequency traps~\cite{cetina12,Joger2014}, i.e. $s$-wave collisions between atoms and ions. For this pair we have $E^*/h \simeq 178.6 \text{ kHz}$, $R^* \simeq 69.8 \text{ nm}$, and $m_\A/m_\I \simeq 0.035.$

Finally, the atom-atom interaction can be treated as short range two-body interaction with lattice EFT~\cite{valiente15}, where the first natural non-zero term affecting spin-polarized fermions is the lowest order odd-wave interaction~\cite{barlette01}. The lattice, with a finite spacing, provides a regularization of the Cheon-Shigehara interaction~\cite{cheon99,valiente15}, and its coupling constant is renormalised by fixing the atom-atom odd-wave scattering length $a_p$.

\subsection{Model atom-ion potential}

The previously introduced polarization potential, \cref{eq:VAI} is state-independent, in that its form does not depend on the internal electronic configuration of the atom and the ion, only the polarisability. However, at short distances, below a few nanometers, the form of the interaction changes to a generally unknown form.

At that spatial range the electronic configurations of the two particles enter into play and render the interaction state-dependent. Such a reliance is included theoretically by assuming that the only effect of the short-range part of the potential on the atom-ion wavefunction is to induce phase shifts. This effect is accounted for by introducing short-range phases $\phi_{e,o}$, which correspond to quantum defect parameters in the context of quantum defect theory~\cite{IdziaszekPRA07,idziaszek11}. Practically, this is handled 
by imposing appropriate boundary conditions in the limit $\vert x - X\vert \rightarrow 0$. In this limit, the polarization potential becomes extremely dominant so that all other energies can be neglected. In 1D such conditions are given by ($X=0$)~\cite{IdziaszekPRA07}
\begin{align}
\psi_e(x) &=  \vert x \vert \sin(1/\vert x \vert + \phi_e) ~,~ x \ll (R^*q)^{-1/2},
\label{eq:evenBound} \\
\psi_o(x) &= x \sin(1/\vert x \vert + \phi_o) ~,~ x \ll (R^*q)^{-1/2},
\label{eq:oddBound}
\end{align} 
with $\psi_{e,o}(x)$ being the even (e) and odd (o) solution of the scattering, respectively, and $q = \sqrt{2 \mu E/\hbar^2}$ with $E$ being the collisional energy at threshold. The short-range phases are free parameters which must be fixed to reproduce the scattering phase shifts found in experiment.
Furthermore, the short-range phases fix the values of the even and odd-wave scattering lengths as
\begin{align}
a_{1D}^{e,o} = R^* \cot(\phi_{e,o}).
\end{align} 
Hence, tuning the short-range phases means to control the above scattering lengths, and therefore the effective 1D atom-ion interaction strength. The above QDT is cumbersome to implement in a many-body Hamiltonian formalism. In order to circumvent this difficulty, we use an effective interaction that faithfully reproduces the long-distance tail of the atom-ion potential, as well as the low-energy phase shifts. In particular, we use the model potential~\cite{schurer14}
\begin{align}
V_{\A \I} (x) = v_0 e^{- \gamma x^2} - \frac{1}{x^4 + 1/\omega},
\end{align}
which is characterised by three parameters: $v_0$, $\gamma$, and $\omega$. We fix $v_0$ at  $3 \omega$ so that the atom wave function (almost) vanishes at $x = 0$, and $\gamma$ is chosen such that
\begin{align}
\gamma \geq \gamma_\text{min} = 4 \sqrt{10 \omega}\,.
\end{align}
In this way, the Gaussian is kept from interfering with the long-range part. 
We can systematically map the free parameter $\omega$ and the semi-restricted parameter $\gamma$ to the quantum defect parameters $(\phi_e,\phi_o)$ (see Appendix A for more details). This means that we can use this potential for numerical modeling, while still considering the quantum defect parameters the tunable parameters of the system.

%
% SECTION
%

\subsection{Atom-atom interaction and discretisation}

We shall solve the many-body problem by discretising it in an equally spaced grid with $N_s$ sites and spacing $d$, giving a total system length of $L = d(N_s-1)$. This will be evaluated in the continuum limit where $L$ remains finite while $d \rightarrow 0$. The discrete Hamiltonian $H_d$ is chosen so that
\begin{align}
\hat{H} = \lim_{d\rightarrow 0} \hat{H}_d.
\label{eq:limiting}
\end{align}
On the lattice (grid), the kinetic part $\hat H_{0,d}$ becomes (as for a Hubbard-like model)
\begin{align}
\hat H_{0,d} = -t(d) \sum_{j=1}^{L-1} \left(\hat c_{j}^\dagger \hat c_{j+1} + \hat c_{j+1}^\dagger \hat c_{j}\right),
\end{align}
where we have in the continuum limit
\begin{align}
t(d) = \frac{\hbar^2}{2 m_\A d^2},
\label{eq:t}
\end{align}
and $\hat c_j (\hat c^\dagger_j)$ is the fermionic annihilation (creation) operator at position $x_j$, respectively.
We will consider the atoms as interacting through van der Waals forces. These can be treated as short range two-body interactions with lattice EFT, where the first natural non-zero term affecting spin-polarized fermions is the lowest order odd-wave interaction ~\cite{valiente15}. In our choice of lattice discretisation this corresponds to a nearest neighbor interaction between the atoms,
\begin{align}
\frac{V_{\text{AA},d}}{t(d)} = \frac{-2}{1 - d/a_p } \sum_{j = 1}^{N_s-1} \hat n_j^\text{A} \hat n_{j+1}^\text{A},
\label{eq:VAA}
\end{align}
where $\hat n^{\A/\I}_j$ is the number operator for atoms/ions. The interaction strength is related to the tunneling rate, the lattice spacing and the p-wave (odd-wave) scattering length $a_p$ via
\begin{align}
V_{\A\A}(d) = \frac{- 2 t(d)}{1 - d/a_p}.
\end{align}
In our calculations we work with $a_p = -0.1 R^*$, corresponding to an attractive interaction without bound states which has strength $V_{\A\A}/t \simeq -1.7$, see Appendix B for details. This particular value was chosen since it gives significant effects while keeping numerical stability. Note that odd-wave interactions may be tuned through e.g. Feshbach resonances or confinement induced resonances~\cite{astrakharchik04,granger04,chin10,zhang04,OlshaniiPRL98,saeidian15}.

To evaluate the ground state of the discrete Hamiltonian we will employ numerical variational calculations using the density matrix renormalisation group (DMRG)~\cite{white92,schoellwock2005}. For such calculations it is convenient to express the Hamiltonian in the characteristic energy $t(d)$, where we combine \cref{eq:units,eq:t} to find the conversion factor
\begin{align}
\frac{E^*}{t(d)} = \frac{(1 + m_\A / m_\I) d^2}{(R^*)^2} = 1.03456 \frac{d^2}{(R^*)^2}.
\end{align}
The effective atom-ion potential is discretised by introducing $x_{ij} = d|i-j|$ and thus the full discretised Hamiltonian is 
\begin{align}
\frac{\hat{H}_d}{t(d)} &= - \sum_{j=1}^{L-1} \left(\hat{c}_{j}^\dagger \hat{c}_{j+1} + \hat{c}_{j+1}^\dagger \hat{c}_{j}\right) +  \frac{-2}{1- d/a_p} \sum_{j = 1}^{N-1} \hat{n}_j^\text{A} \hat{n}_{j+1}^\text{A} \nonumber\\
\phantom{=} &+ \frac{E^*}{t(d)} \sum_{i,j} \hat{n}^\text{I}_{i} \hat{n}^\text{A}_{j} \left( v_0 e^{-\gamma x_{ij}^2} - \frac{1}{x_{ij}^4 + 1/\omega} \right),
\label{eq:simHam}
\end{align}
which satisfies \cref{eq:limiting} up to a constant energy shift.

\begin{figure}[tb]
\centering
\includegraphics[width=\columnwidth]{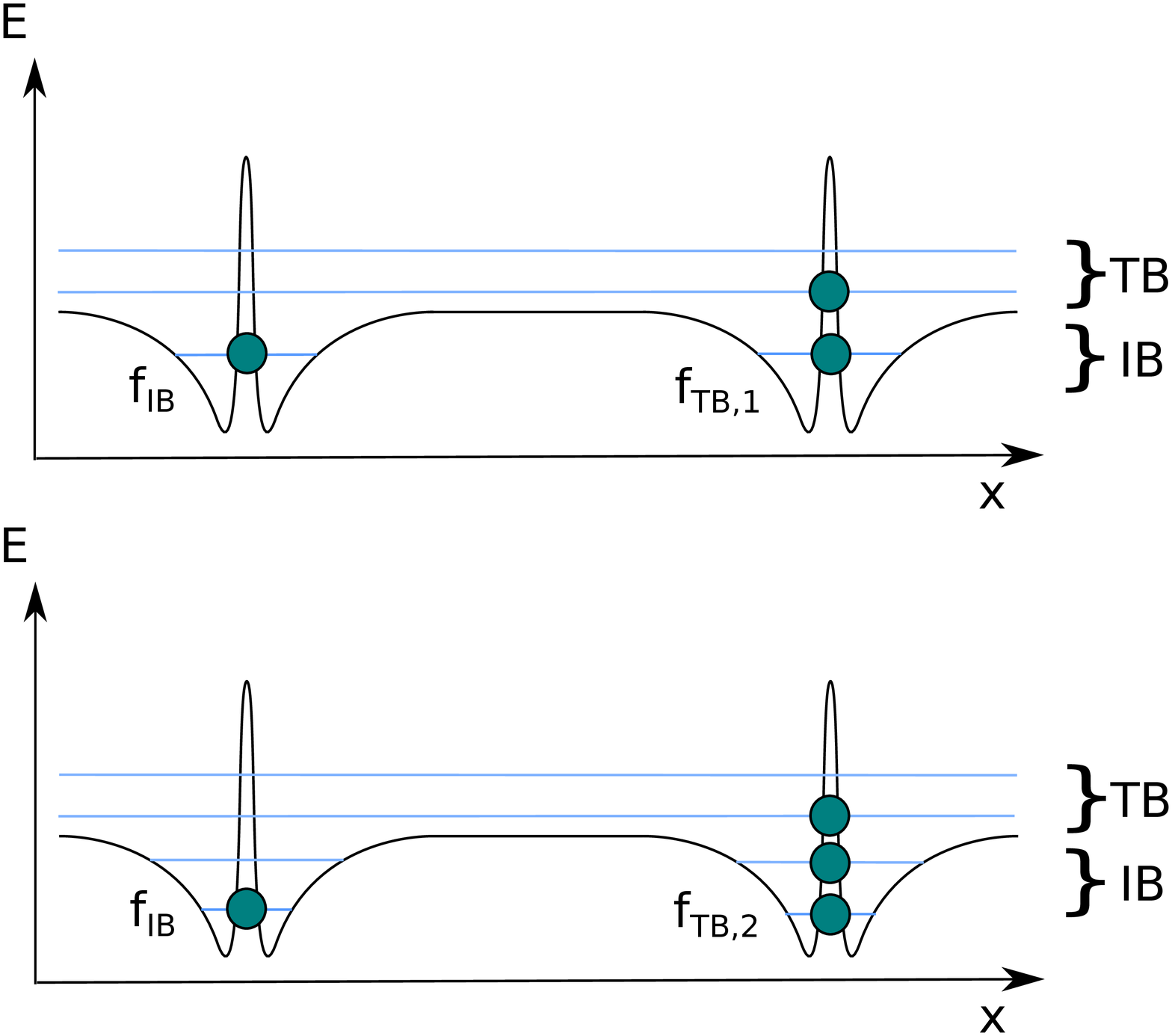}
\caption{A diagram of the effective potential generated by two ions. The black line shows the total potential, in this case a box with two ions as described by the model potential, whereas the blue horizontal lines indicate the energy of a specific eigenstate. A blue circle on a blue line indicates that this eigenstate has been occupied by a fermion. The number of ion-bound states per ion depends on the model parameters. The upper diagram shows the ion-bound (IB) and trap-bound (TB) filling types for the case of one ion-bound state, the lower for the case of two ion-bound states. }
\label{fig:energylevels2}
\end{figure}

For the range of QDT parameters we investigate, the atom-ion interaction supports one or two two-body bound states. For ions in a finite lattice with open boundaries, this means we have two type of states (see \cref{fig:energylevels2}): states deep in the effective atom-ion potential which would not exist in a flat potential, corresponding to ion-bound (IB) atoms, and a discrete set of states above the IB states similar to those found in a 1D quantum well, which we will call trap-bound (TB) since the discretisation is due to the presence of the (box-like) trap. Note that the TB states are still affected by the presence of the ions. We will consider two different $N_A/N_I$ fillings of the system. An $f_\text{IB}$ filling, where $N_\A = N_\I$, and each atom will occupy an IB state, and two $f_\text{TB}$ fillings, where all IB states are filled and $N_\I$ atoms are added, which, because of quantum statistics, occupy $N_\I$ TB states. Two such fillings must be considered to take into account the difference in the number of IB states.

%
% SECTION
%

\subsection{Tomonaga-Luttinger liquid theory}
\label{sec:TLL}
\begin{figure}[tb]
\centering
\includegraphics[width=\columnwidth]{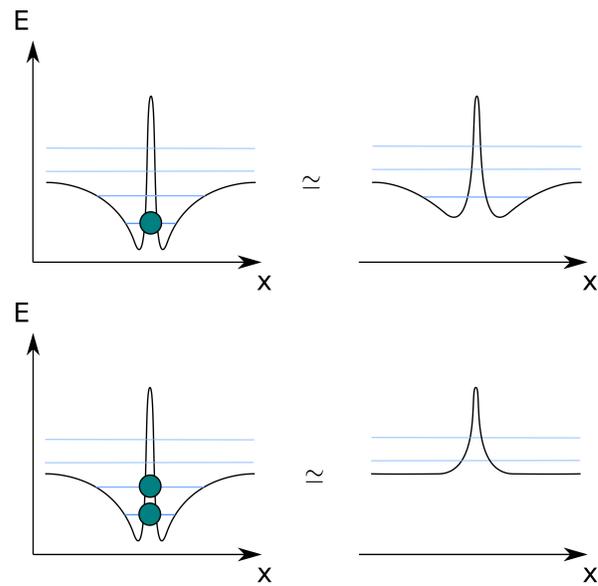}
\caption{Filling of the lower energy states will correspond to a screening of the attractive part of the atom-ion potential. As shown on this diagram, this can be understood microscopically as an effective cancellation of the wells on either side of the ion, ultimately only leaving a soft barrier. The expected effect in TLL terms would be a raising of the value of $K$ for low filling, corresponding to induced attraction, and a lowering of the value of $K$ for high fillings, corresponding to induced repulsion.}
\label{fig:screening}
\end{figure}

A system of interacting fermions in one dimension is fully characterised at low energy by the renormalized speed of sound $u$ and the TLL parameter $K$, which is a dimensionless parameter with $K < 1$ for repulsive fermions, $K = 1$ for non-interacting fermions, and $K > 1$ for attractive fermions. The goal of our study is to investigate the impact of an ion lattice on such parameters. From the previous discussion on fillings we would expect the ions to act as attractive wells in the low filling cases, $f_\text{IB}$. As the filling rises, the attraction becomes screened by the atoms, and at high filling, $f_\text{TB}$, the atom-ion potentials effectively become soft barriers as shown pictorially in \cref{fig:screening}. The expected effects are that in the $f_\text{IB}$ case, the atoms are forced closer together, effectively increasing their mutual attraction, i.e. the value of $K$ would rise. In the $f_\text{TB}$ case, the atom-ion potential is repulsive, and the expected effect is an induced repulsion between atoms, corresponding to a lowering of the value of $K$. In all cases we would expect a lowering of the speed of sound due to the introduction of barriers in the fluid, corresponding to a higher effective atomic mass. However, none of these behaviors follow trivially from the shape of the potential. Note that the degree of all these effects will depend on the nature of the atom-ion interaction as determined through the short-range phases $\phi_{e,o}$.

The ground state properties of the quantum fluid can be analyzed through the bosonized Hamiltonian
\begin{align}
\hat H &= \frac{1}{2 \pi} \int \left[u K (\partial_x \theta)(x)^2 + \frac{u}{K}(\partial_x \phi)(x)^2 \right] dx,
\label{eq:LLHamil}
\end{align}
where $\theta$ and $\phi$ are the standard bosonic fields. 
This effective Hamiltonian is a linearisation of \cref{eq:Hmicro} around the Fermi points. We will extract $u$ and $K$ as functions of the quantum defect parameters, by treating the microscopic Hamiltonian, \cref{eq:Hmicro}, in a DMRG calculation and evaluating the ground state properties of systems with varying quantum defect parameters. This will allow us to extract the TLL parameters using the methods outlined below. Let us stress here that the discretisation has no physical significance, but it is done merely to allow a numerical treatment of the continuous system.

Specifically, we consider that on the $N_s$ sites of our system there are $N_\A$ atoms and $N_\I$ ions. When $N_\A, N_\I \ll N_s$ and $d \ll R^*$ (i.e. low filling factor), we can use DMRG on the discretised system to approximate the $d \rightarrow 0$ continuum limit~\cite{bellotti17,dehkharghani17}. When we approach the thermodynamic limit numerically
\begin{align}
N_s \rightarrow \infty ~~,~~ d = \text{const.},~~ N_{\A/\I}/N_s = \text{const.},
\end{align}
we can extract $K$ from the momentum space density-density correlation function for the minimum lattice momentum $k_0 = 2 \pi/N_s$ as~\cite{ejima05}
\begin{align}
K = \lim_{N_s \rightarrow \infty} 2 \Big( \big\langle \hat n\left(k_0 \right) \hat n\left(-k_0\right) \big\rangle - \big\langle \hat n(k_0) \big\rangle \big\langle \hat n(k_0) \big\rangle \Big).
\label{eq:K}
\end{align}
Here the expectation value is with respect to the ground state $\psi_0$ of the fermionic system. We have used the Fourier transformed number operator
\begin{align}
\hat n(k) = \hat n^\dagger(-k) = \sum_{j=1}^{N_s} e^{-ik(j-j_c)} \hat c_j^\dagger \hat c_j,
\end{align}
with $k$ being lattice momentum, $j$ being the lattice site index and $j_c$ being the central site. To reach this limit we use \cref{eq:K} on a number of finite systems with increasing size and constant lattice spacing, atom density and ion density. We then extrapolate $K$ to the infinite size limit using a linear fit, see Appendix B for further details. 

In order to find $u$ we estimate the compressibility $\kappa$ of the system, whose inverse is related to TLL theory, \cref{eq:K}, as~\cite{giamarchi03_3}
\begin{align}
\frac{1}{\kappa} &= \frac{u \pi}{K} = \frac{L}{2} \dd{E}{N_\A} \nonumber\\
\phantom{=}&\simeq \frac{L}{2} \left( \frac{E(N_\A+2) + E(N_\A-2) - 2E(N_\A)}{4} \right),
\label{eq:kappa}
\end{align}
where $E(N_A)$ is the energy of a system with $N_A$ atoms. The factor $1/2$ in the second line of \cref{eq:kappa} accounts for the spin polarization. The derivative must be approximated as a finite difference, since the number of particles is discrete, and we use a difference of two atoms to avoid any effects which might arise due to the differences between having an odd and an even number of particles. By computing the ground state energy of the system for different numbers $N_A$ of fermions, we can thus calculate both TLL parameters by using \cref{eq:kappa,eq:K}.

Since the ions in our systems are equally spaced, effectively forming a periodic potential, the non-interacting variant can be accurately described using Bloch waves and band theory~\cite{NegrettiPRB14} in the thermodynamic limit. Such a system contains gaps between the bands at integer fillings, i.e. $N_\A = n N_\I$ where $n$ is an integer. If our system is in such a gapped state it cannot be modelled using TLL theory. However, we are considering a system of interacting atoms, where the lattice model of the atom-atom interaction \cref{eq:VAA} is inversely proportional to the lattice spacing. By approximating the continuum with a small lattice spacing, the interaction becomes comparably large, which can lead to a closing of said gaps, and ensure non-insulating behavior. However, an interacting system might still be a Mott insulator. To classify the behavior of the systems treated, we have calculated $\kappa$ for each system and extrapolated it to the thermodynamic limit. In this limit, $\kappa \rightarrow 0$ for any type of insulator since the energy gap causes the energy difference in \cref{eq:kappa} to remain finite at infinite length. It was found that none of the systems treated exhibited such behavior, with the smallest extrapolated value being $\kappa = 0.16(3)(R^*E^*)^{-1}$. From this we conclude that all systems considered can be accurately modelled using TLL theory.

In the rest of the paper we will assume an ion density of $N_\I/L = 0.25/R^*$. This means that in the thermodynamic limit, the ions have a separation of $D = 4 R^*$, which for the atom-ion pair $^6\text{Li}$ / $^{174}\text{Yb}^+$ corresponds to 279.2 nm. In the case of $N_\I = 7$, the ion spacing is $D=4.6 R^*$, corresponding to 321.1 nm. For the atom-ion pair $^{40}\text{K}$ / $^{174}\text{Yb}^+$ the ion spacing would correspond to 1.1 $\mu$m. For instance, for a $^{174}\text{Yb}^+$ ion chain with $N_\I = 7$ ions and a radiofrequency of 2$\pi\times$ 2 MHz, the minimal separation is about 1.2 $\si{\micro\meter}$, whereas with a radiofrequency of 2$\pi\times$ 10 MHz it is 398.22 nm~\cite{James1998}. Although the latter frequency is higher than typical values encountered in experiments, the quoted separations can be obtained by just generating time-dependent fields of higher frequency. Attempts at reaching ion separations that are currently attained in trapped ion experiments is beyond the capabilities of our DMRG calculations. Nonetheless, since the smaller ion separation we have considered, i.e. $D = 4 R^*$, is large enough that the atom-ion potentials have negligible overlap (see also Fig.~\ref{fig:energylevels2}), we do not expect any qualitative differences from increasing the separation.

%
% SECTION
%

\section{Results}

The following results were obtained by using the DMRG algorithm as outlined above. Errors on $K$ are the $2 \sigma$ confidence intervals in the linear fits used for extrapolation. To ensure the correct implementation of our method we tested the calculation without ions. For $a_p=0$ we find the free fermion limit $K = 1.0000(2)$, as expected for non-interacting atoms, while the slightly attractive interaction $a_p = -0.1 R^*$ gives $K=1.0525(9)$. This is similar to the result we get by approximating the fermions as hard rods~\cite{mazzanti08} with length $a_p$ in a system with fermionic density $\rho$, $K_\text{hs} = (1 - \rho a_p)^2 = 1.0506$. When comparing the calculated speed of sound for a free fermion gas with the Fermi velocity $v_F$ of the same system, we find $u/v_F = 1.03(5)$, where the error is due to discretisation.

The parameter space which gave significant effects while being numerically feasible was found to be~\cite{michelsen2018} 
\begin{align}
\label{eq:range}
1 \leq \frac{\omega}{(R^*)^{-4}}, \frac{\gamma}{\gamma_\text{min}} \leq 10,
\end{align}
where the combinations
\begin{align}
\label{eq:omega-gamma}
\frac{\omega}{(R^*)^{-4}} = 2, 4, 6, 8, 10 ~~,~~ \frac{\gamma}{\gamma_\text{min}} = 1, 2, 5, 10
\end{align}
give a relatively even spread of quantum defect parameters. Importantly, there is a transition in the number of bound states per ion within this parameter space, see \cref{tab:bound_states}. In the rest of the text the systems with two bound states will be said to be in the ``strong" ion domain (since the potential has deeper wells and higher central Gaussian), while the systems with one bound state will be said to be in the ``weak" ion domain. In the QDT parameter plots, \cref{fig:KmapIBa,fig:umapIBa,fig:KmapTBaWeak,fig:KmapTBaStrong}, this transition is schematically marked with a dashed line.
\begin{table}[b]
\caption{Number of ion-bound atomic states per ion for those of the model parameter combinations considered in this study involved in the transition from one to two such states. This transition is marked schematically by a dashed line in \cref{fig:KmapIBa,fig:umapIBa,fig:KmapTBaWeak,fig:KmapTBaStrong}}
\begin{tabular}{c|c|cccc}
\multicolumn{1}{r}{} & \multicolumn{1}{r}{} &\multicolumn{4}{c}{$\gamma/\gamma_\text{min}$}\\
\cline{3-6}
\multicolumn{1}{r}{} & & 1 & 2 & 5 & 10\\ 
\cline{2-6}
\multirow{3}{*}{$\omega/(R^*)^{-4}$} &4 & 1 & 1 & 1 & 1\\ 
&6 & 1 & 2 & 2 & 2\\ 
&8 & 2 & 2 & 2 & 2\\ 
\end{tabular} 
\label{tab:bound_states}
\end{table}
This is particularly relevant for the investigation of TB states, and will be discussed further in \cref{sec:TB} below. Note that ions located on an edge site will always have one bound state, which is a finite size effect. Further technical details can be found in Appendix B. Finally, for all plots in the following section, the points signify calculation results and the surface is a linear interpolation.

%
% SECTION
%

\subsection{Ion-bound atomic states}
\begin{figure}[tb]
\centering
\includegraphics[width=0.9\columnwidth]{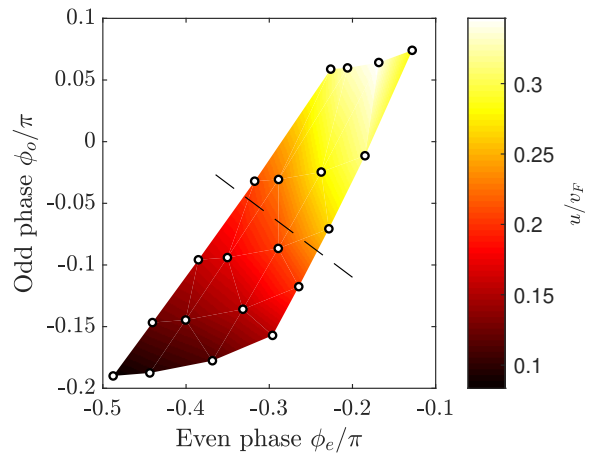}
\caption{Speed of sound for ion-bound states of a system of ions and interacting fermionic atoms with $N_A = N_I$. All points have errors of $\pm 0.05$. The dashed line schematically marks the transition from the weak ion domain (above the dashed line) the strong ion domain (below the dashed line), see the text for further details. The ions significantly lower the speed, corresponding to a hindering of collective excitations, especially in the strong domain (below the dashed line).}
\label{fig:umapIBa}
\end{figure}
\begin{figure}[tb]
\centering
\includegraphics[width=0.9\columnwidth]{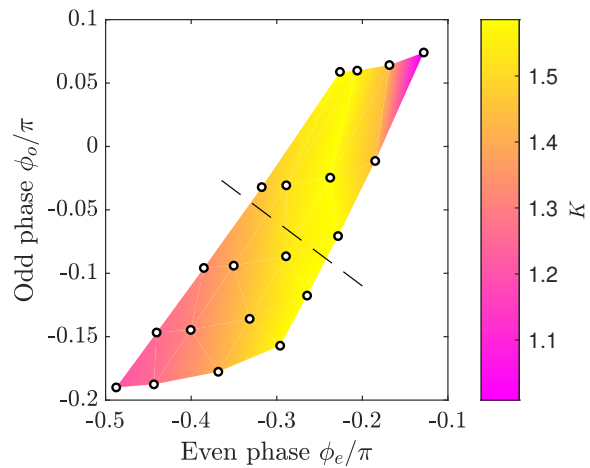}
\caption{TLL parameter for a range of QDT parameters in systems of ions and interacting fermionic atoms with filling $N_\A = N_\I$. All results have errors less than $ \pm 0.02$.}
\label{fig:KmapIBa}
\end{figure}

In \cref{fig:umapIBa} we show the speed of sound $u$ of the system of interacting atoms and ions. Generally, the presence of the ions lowers this speed considerably compared to the Fermi velocity of a free fermion system, with the clearest effects in the strong domain. Since we are effectively introducing potential wells and barriers into the system, it is to be expected that collective excitations across the systems will be damped by these ``obstacles", corresponding to a lower speed of sound, or equivalently a higher effective mass of the fermions.
The introduction of ions into our system of interacting fermions induces a significant effective attraction between the interacting atoms, as shown in \cref{fig:KmapIBa}, where the $K$-parameter varies approximatively from 1.20 to 1.58, with a dip to 1 in the deep weak domain (above the dashed line). This depends mostly on $\phi_e$, and peaks for $-0.3 < \phi_e/\pi < -0.2$. For values larger than this, we see hints at a sharp dip towards the non-interacting limit.

%
% SECTION
%

\subsection{Trap-bound atomic states}
\label{sec:TB}

\begin{figure}[tb]
\centering
\includegraphics[width=0.9\columnwidth]{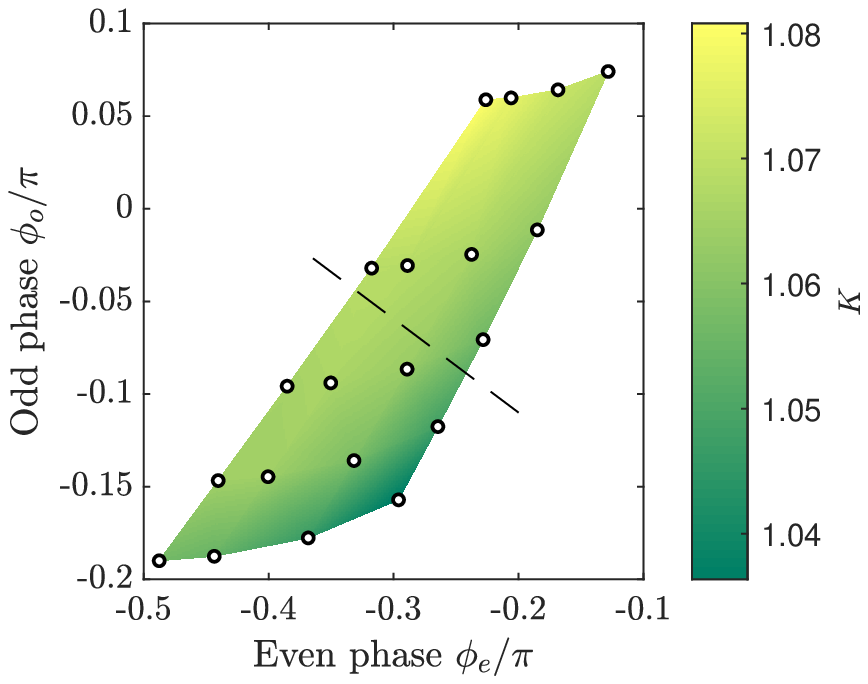}
\caption{Luttinger liquid parameter for a range of QDT parameters in systems of ions and interacting fermionic atoms with filling $N_\A = 2 N_\I$. All results have errors less than $\pm 0.02$.}
\label{fig:KmapTBaWeak}
\end{figure}

\begin{figure}[tb]
\centering
\includegraphics[width=0.9\columnwidth]{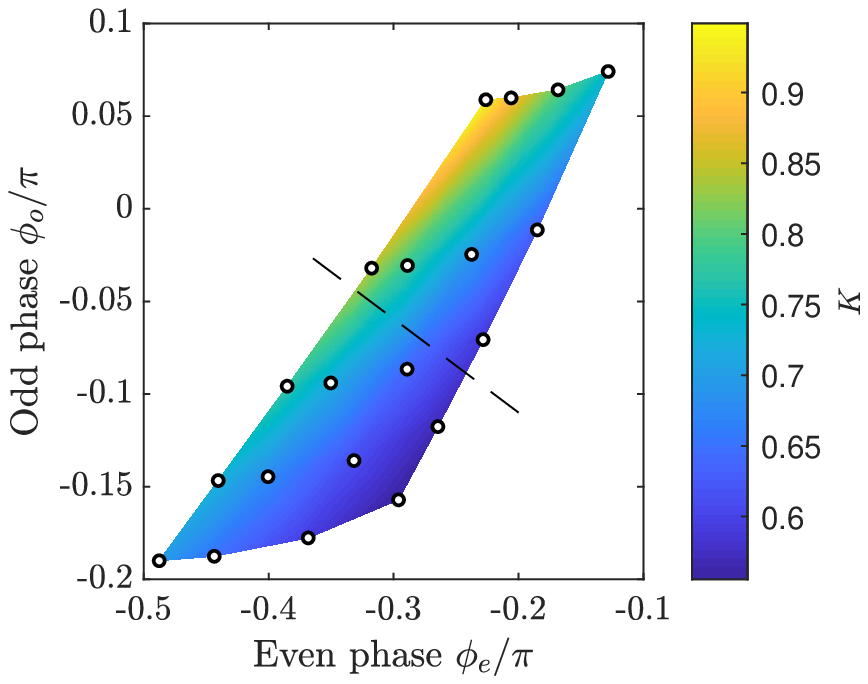}
\caption{Luttinger liquid parameter for a range of QDT parameters in systems of ions and interacting fermionic atoms with filling $N_\A = 3 N_\I - 2$. All results have errors less than $\pm 0.05$.}
\label{fig:KmapTBaStrong}
\end{figure}

Due to the previously mentioned transition in the number of IB states per ion, in order to study the behavior of a system of TB states we must consider different fillings in the different domains. In the weak domain we consider the $f_\text{TB,1}$ filling $N_\A = 2 N_\I$, while in the strong domain we consider the $f_\text{TB,2}$ filling $N_\A = 3 N_\I-2$, where two states are subtracted due to the fact that the ions at the edges can only host one odd-wave bound state. \cref{fig:KmapTBaWeak} shows the $f_\text{TB,1}$ filling over both domains, and with $K$ varying between approximatively between 1.03 and 1.08, we see that the ions barely tune it away from the $1.05$ value from the system with no ions, with slightly induced attraction in the weak domain. \cref{fig:KmapTBaStrong} shows the $f_\text{TB,2}$ filling over both domains, and with $K$ varying from 0.56 to 0.95 we can see a strong induced repulsion. Remarkably, there is a smooth transition between domains for both fillings, but drastically different $K$-values between the fillings, suggesting that the deciding factor in the value of $K$ is not the density of TB or IB states, but rather the total number of atoms per ion. The smooth transition between domains indicate that the difference between a low TB state and a shallow IB state has very little influence on the physics of our system. 

\subsection{Discussion}

Taken together, \cref{fig:KmapIBa,fig:KmapTBaWeak,fig:KmapTBaStrong} indicate that the effect of the ions on the atom-atom interaction can be separated into three different categories
\begin{enumerate}
\item \emph{Stronger attraction.} This is the case when \mbox{$N_\A = N_\I$}, and the atoms are bound relatively deep in the atom-ion potential.
\item \emph{No effect or slightly stronger attraction}. This is the case when \mbox{$N_\A = 2 N_\I$}, with slightly more attraction in the weak domain.
\item \emph{Shift to repulsive interaction.} This is the case when \mbox{$N_\A = 3 N_\I-2$} in the strong ion domain, with tendency towards the non-interacting limit in the weak domain.
\end{enumerate}
As predicted in \cref{sec:TLL}, we have induced attraction at low fillings (cat. 1), which transitions over a cancellation of attractive and repulsive effect (cat. 2) to induced repulsion at high fillings (cat. 3). A remarkable result is the shift of $K$ from $K > 1$ to $K < 1$, meaning that the introduction of ions causes the initially attractive atoms to have an effectively repulsive interaction. These results are in clear contrast to the behavior of a TLL in a flat potential, where changing the atomic density cannot tune the effective interaction across the free fermion limit~\cite{giamarchi03_5}. To tune an initially attractive TLL into a repulsive TLL through a change in the atom density thus requires an inhomogeneous potential such as the one generated by an ion chain.

The fact that the ions induce repulsion between the fermions indicates that the atomic gas has a tendency to form a so-called charge density wave, i.e. an ordered state. In our setting this means a density wave of fermionic atoms. A similar phenomenon has been observed for a 1D Fermi gas coupled parallel to an ion chain~\cite{BissbortPRL13, giamarchi03_4}, where the (transverse) atom-phonon coupling induces a Peierls instability below a critical separation between the two quantum systems.

Current experiments with ytterbium ions and lithium atoms~\cite{Furst2017} show very low Langevin collisional rates, thus indicating that atoms do not occupy bound states within the ions, and so the points 2 and 3 above are the most experimentally relevant. Note that these effects are genuinely induced by the atom-ion scattering physics, that is, the occurrence of one or two bound states at threshold is a physical effect tuneable by Feshbach or confinement-induced resonances.

%
% SECTION
%

\section{Conclusions and outlooks}

We have investigated the ground state properties of a fermionic quantum fluid superimposed on a uniform ion chain. Particularly, we have assessed the Luttinger liquid parameters $K$ and $u$, which fully characterise the ground state of the spin-polarised Fermi gas and its low-energy excitations. Our goal was to analyse the reliance of the TLL parameters on the short-range phases of the atom-ion scattering. To this aim, we performed numerical density matrix renormalisation group calculations on a high-resolution discretised fermionic Hamiltonian modeling a static linear ion chain. Thus, we have been able to map the Luttinger liquid parameters to the two short-range phases characterising the atom-ion polarization potential. By changing these scattering parameters, e.g. via external driving of the ionic impurities, we have shown that the Luttinger liquid parameters can be tuned within a broad range of values. While the speed of sound is generally decreased, corresponding to a hindering of collective excitations by the ions, the interaction as measured by $K$ has a more intricate behavior. Depending on the density of the initially weakly attractive atoms, changing the ion scattering parameters can tune the interaction within a repulsive regime, an attractive regime or have completely negligible effect. The result of most immediate experimental relevance is the induced repulsion.

Finally, future work could address the dimensional crossover by replacing the setup we investigated purely in 1D with an atomic waveguide, where the motional transverse degrees of freedom are taken into account, too. Recently the analytical solution of the 3D scattering problem of a trapped atom interacting with an array of contact potentials, i.e. representing the static scattering centres akin to the ions, was presented~\cite{MartaPRA18}. Hence, one could solve the many-particle problem using this analytical solution and investigate the impact of the transverse confinement of the atoms on the TLL parameters and excitation spectrum of the liquid in order to understand the interplay between external confinement and impurity-atom scattering characteristics. An alternative approach could be by means of bosonisation techniques, where the transverse modes are coupled~\cite{Kamar2019}. Moreover, another interesting research direction is to study the role of spatial inhomogeneities in the impurity-atom interaction strength, thus adding controlled disorder in the system.

%
% SECTION
%

\section{Acknowledgements}

ABM would like to thank Rafael Barfknecht and Satoshi Ejima for fruitful discussions. This work was supported by the Aarhus University Research Foundation under a JCS Junior Fellowship and the Carlsberg Foundation through a Carlsberg Distinguished Associate 
Professorship grant, by the Cluster of Excellence projects ``The Hamburg Centre for Ultrafast Imaging" of the Deutsche Forschungsgemeinschaft (EXC 1074, Project No. 194651731) and ``CUI: Advanced Imaging of Matter" of the Deutsche Forschungsgemeinschaft (EXC 2056, Project No. 390715994), and the National Research Fund, Luxembourg, under grant ATTRACT 7556175. The numerical work presented in this paper was partially carried out using the HPC facilities of the University of Luxembourg~\cite{VBCG_HPCS14}
{\small -- see \url{https://hpc.uni.lu}}.

%
% SECTION
%

\appendix
\setcounter{figure}{0}
\renewcommand{\thefigure}{A\arabic{figure}}

\section{The model potential}
For the sake of numerical efficiency, we have chosen the atom-ion model potential parameters within the range
\begin{align}
1 \leq \frac{\omega}{(R^*)^{-4}}, \frac{\gamma}{\gamma_\text{min}} \leq 10.
\end{align}
The mapping between the QDT parameters, i.e. short-range phases, and model potential parameters is performed by following this procedure:
\begin{enumerate}
\item We choose some values for $\omega$ and $\gamma$ within the range outlined above as well as $E = k^2$ (all parameters are in units of $E^*$ and $R^*$), the latter of which must be small (i.e. in the low-energy limit), but positive. We then use the Numerov method~\cite{johnson77} to solve the Schr\"odinger equation for the two-body atom-ion problem for this potential by iterating the wavefunction from $x = 0$ to $x \gg R^*$.
\item We determine the phase shifts $\xi_{e,o}$ of the solution at large distances, i.e. far from the ion, by comparing the logarithmic derivative of the solution to a plane wave solution at $x = x_0 \gg R^*$,
\begin{align}
\cot(\xi_{e,o}) &= \frac{k + A_{e,o} \cot(k x_0)}{-A_{e,o} + k \cot(k x_0)}, \\
A_{e,o} &= \left.\frac{d\psi_{e,o}(x)/dx}{\psi_{e,o}(x)}\right\vert_{x=x_0}.
\end{align}
\item We test QDT solutions of different $\phi_{e,o}$ and determine the corresponding phase shifts as in the previous step 2.
\item We compare the phase shift, $\xi_{e,o} (\phi_{e,o})$, obtained via QDT for a certain pair of short-range phases $\phi_{e,o}$ with the sample of phase shifts, $\xi_{e,o} (\omega, \gamma)$, obtained with the model potential for various parameters $\omega, \gamma$. The one that is most similar to $\xi_{e,o} (\phi_{e,o})$ gives the mapping.
\end{enumerate}
We note that the last step of this procedure always yields a numerical error, i.e. the difference between the QDT result and the model potential will be around $10^{-12}$. We also note that for perfect precision in the mapping, the atom-ion wave-function would have to be zero at the ion position. This is only true for the model potential to a good approximation, since the model parameter $v_0$ is finite.

%
% SECTION
%

\section{DMRG calculations and extrapolation}
\setcounter{figure}{0}
\renewcommand{\thefigure}{B\arabic{figure}}
\begin{figure}[tb]
\centering
\includegraphics[width=0.9\columnwidth]{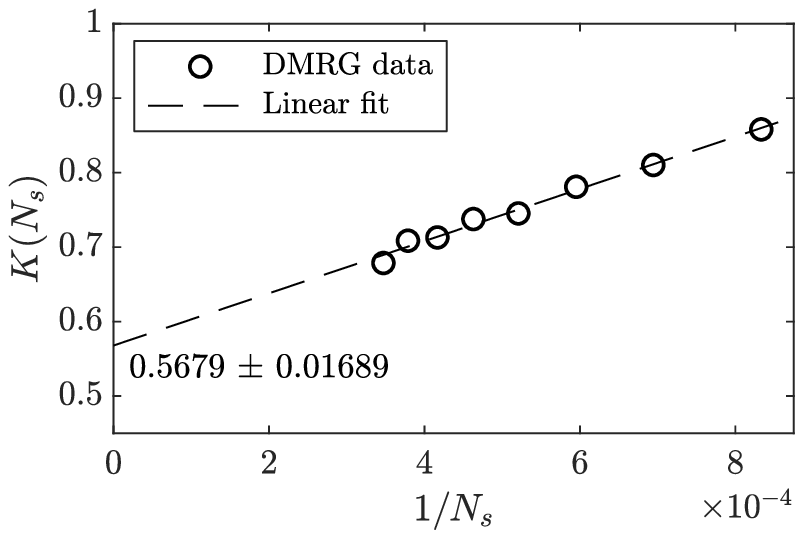}
\caption{An example of values of $K(N_s)$ for trap-bound systems with constant atom and ion densities, constant lattice separation and different sizes (as measured by number of lattice points $N_s$). To calculate $K = \lim_{N_s \rightarrow \infty} K(N_s)$ we apply  to finite systems and extrapolate the results to the infinite size limit $1/N_s \rightarrow 0$ using a linear fit. The value of $K$ in the limit is shown with an error which is the $2 \sigma$ confidence interval on the fit. This example has \mbox{$\omega = 10 (R^*)^{-4}$}, $\gamma = \gamma_\text{min}$ and $a = 0.1 R^*$ and $N_\A = 3 N_\I - 2$.}
\label{fig:extrp}
\end{figure}

The DMRG solutions were found by using the implementation from the \textsc{ITensor} library~\cite{miles18}.
The time taken and accuracy achieved depends on a number of supplied parameters:
\begin{description}
\item[Sweeps] The number of sweeps to achieve convergence depends heavily on the size and complexity of the system of interest, ranging from $\sim 100$ for small atom-only systems with simple interactions, to 1000-2000 for large atom-ion systems with many atoms and all interactions turned on.
\item[Cutoff] DMRG uses a singular value decomposition (SVD) procedure, where all singular values below this cutoff value are truncated. The value was kept similar to that of Refs.~\cite{dehkharghani17, bellotti17}, namely $\sim 10^{-13}$.
\item[Maximum bond dimension] It was found that setting this value at $1000$ gives a good convergence time.
\end{description}

The \textsc{ITensor} implementation automatically converts common operators into matrix product operators (MPOs). This renders the implementation of the Hamiltonian as well as the extraction of the ground state energy and the density profile
\begin{align}
\expect{\hat{\rho}(x_j)} = \bra{\psi_0} \frac{\hat{c}^\dagger_j\hat{c}_j}{d} \ket{\psi_0}.
\end{align}
rather simple.
A straightforward way to confirm that the algorithm has converged is to check the symmetry of the density profile. The true ground state will be completely symmetric around the center of the trap, but it was found that the DMRG algorithm would only return states with symmetric density profiles once it had completely converged. 

The parameter space which gave significant effects while being numerically feasible was found to be~(\ref{eq:range}), 
whereas the combinations~(\ref{eq:omega-gamma})
give a nice spread of quantum defect parameters. Smaller parameters would make the features of the potential too weak, while larger parameters tend to give an non-smooth potential, requiring a finer lattice to properly resolve. Within this parameter range it was found that a lattice constant of $d \sim 0.01 R^*$ with $\sim 400$ sites per ion was a minimum for reliable calculations. 
Extrapolation was done from the results of calculations with 5 to 12 ions, a density of $N_\I/L = 0.25/R^*$ and a lattice separation of $d = 0.01667 R^*$, see \cref{fig:extrp}. Since the data points cluster closer together towards $1/N_s \rightarrow 0$, and to have more efficient calculations, it was chosen to only extrapolate using $N_\I = 5,6,7,9,12$, which still gives a reliable extrapolation. For $\omega/(R^*)^{-4} < 5$ the $N_I=5$ results were found to be unreliable and had to be excluded from the extrapolation. The remaining points sufficed for reliable extrapolation.

The main system of interest in this paper is that with trap-bound filling and non-zero atom-atom interactions. It is however noteworthy that the extrapolation procedure failed for the ion-bound filling when atom-atom interactions were neglected (i.e. $V_{\A\A} = 0$). One would expect this to be a simpler system to work with, but our numerical procedure failed in this case.
Extraction of $K$ using \cref{eq:K} can be readily done by seeing that the sum is symmetric around $j_c$, meaning the imaginary parts of the exponential cancel, and one is left with
\begin{align}
\hat n(k) = \hat n(-k) = \sum_{j=1}^{N_s} \cos[ k (j-j_c) ] \hat c^\dagger_j \hat c_j,
\end{align}
which is real and even, and which can be converted to an MPO and applied to the ground state before calculating the overlap in  \cref{eq:K}.

%
% SECTION
%

%-------------------------------
\bibliography{zotero}
%-------------------------------

\end{document}